\documentclass[12pt,epsf,showkeys]{article}

\usepackage{graphicx,float}
\usepackage{mathrsfs,array,multirow}
\usepackage{amstext}
\usepackage{subfigure}
\usepackage{bm,latexsym,amsmath,amsfonts,amssymb}
\usepackage[usenames,dvipsnames]{color}
\usepackage{color}
\usepackage[colorlinks=true,linkcolor=blue]{hyperref}
\usepackage{soul}
\usepackage{epsfig}
\usepackage{subfigure}	

\newcommand{\be}{\begin{equation}}

\newcommand{\ee}{\end{equation}}

\newcommand{\ba}{\begin{array}}

\newcommand{\ea}{\end{array}}

\begin{document}
\begin{titlepage}
\vspace{.5in}

\begin{center}
{\Large\bf Anisotropic matter and nonlinear electromagnetics \\ black holes }\\
\vspace{.4in}
  {$\mbox{Wonwoo Lee}$}\footnote{\it email: warrior@sogang.ac.kr} and
  $\mbox{Yun Soo Myung}$\footnote{\it email: ysmyung@inje.ac.kr}\\

\vspace{.3in}

{ Center for Quantum Spacetime, Sogang University, Seoul 04107, Korea}\\

\vspace{.5in}
\end{center}
\begin{center}
{\large\bf Abstract}
\end{center}
\begin{center}
\begin{minipage}{4.75in}

It is shown that anisotropic matter black holes with two parameters $w$ and $K$ correspond to nonlinear electrodynamics (NED) black holes with power-index $s$ and  charge term $\xi(s,q)$ by introducing a NED term.
These NED black holes include dark matter ($s=3/4$), constant scalar hair ($s=1$), charged quantum Oppenheimer-Snyder ($s=3/2$), and  Einstein-Euler-Heisenberg ($s=2$) black holes derived from their known actions.
Rotating  NED black holes can be obtained from rotating anisotropic matter  black holes when replacing $w$ and $K$ by $2s-1$ and $\xi(s,q)$. The extremal rotating NED
black holes being  the boundary between rotating charged NED black hole and naked singularity are derived as functions of the rotation parameter $a(q)$.

\end{minipage}
\end{center}
\end{titlepage}

\newpage
\section{ Introduction \label{sec1}}

Recent observational reports regarding cosmic expansion and galactic supermassive black holes would appear to challenge the existing paradigm of our Universe~\cite{DESI:2025zgx, Son:2025rdz, Capozziello:2025qmh, EventHorizonTelescope:2025vum, EventHorizonTelescope:2025whi, EventHorizonTelescope:2025dua}, thus prompting a reassessment of our fundamental cosmological models. The current cosmological model suggests that dark energy and dark matter constitute most of the total energy budget, while ordinary matter only accounts for about a few percents~\cite{Planck:2018nkj}. This realization of our limited knowledge underscores the necessity for a more cautious and humble approach to exploring the Universe.

The majority of compact objects would be rotating in the Universe, and there has been a concomitant increase in interest in solutions to Einstein's equations for rotating black holes~\cite{Kerr:1963ud, Newman:1965my}. Furthermore, supermassive black holes have been found at the centers of galaxies~\cite{Ghez:2008ms, Gillessen:2008qv}. These black holes coexist with dark matter in astrophysical environments and are becoming a focal point of study~\cite{Fernandes:2025osu}. Accordingly, research interests are increasingly shifting towards black hole solutions that incorporate matter fields, moving beyond Einstein's vacuum solutions~\cite{Kim:2026gog, Datta:2026krm, Konoplya:2025ect, Podolsky:2025tle, Ovcharenko:2025fxg,  Kim:2025sdj}.

Anisotropic matter black holes (AMBHs) with equation of state (anisotropic)  parameter $w$ and energy density  $K$ were derived from an anisotropic fluid with equation of states ($p_1=-\rho, p_2=w\rho$)\cite{Cho:2017nhx}. Here, $w$ and $K$ may be regarded as hairs of AMBH.
It is worth to note that AMBHs and their  charged solution  were also found from the quintessential energy-momentum tensor $T^q_{\mu\nu}$~\cite{Kiselev:2002dx, Toshmatov:2015npp}.
We note that the power-law hairy black hole solution was found from the Einstein-Kalb-Ramond theory~\cite{Kumar:2020hgm}.
Furthermore, rotating charged AMBH solution  were derived by making use of the Newman-Janis algorithm~\cite{Kim:2019hfp}.
This  solution is considered as an extension of the Kerr-Newman (KN) black hole.
Its shadow analysis was performed  by discussing the effect of the anisotropic term ($K/r^{2w-2}$ )~\cite{Lee:2021sws}.
On the other hand, the anisotropic (exotic) fluid could  be used to construct the Morris-Thorne-type wormhole under the  condition of  $p_1+2p_2=-\rho$~\cite{Morris:1988cz}.

So far, there is no known action corresponding to this anisotropic fluid and thus it may  restrict researchers from a further study on AMBHs.

In the present work, we introduce the nonlinear electrodynamics (NED) term of $\mathcal{F}^s$ with $\mathcal{F}$ the Maxwell term~\cite{Hassaine:2008pw, Dariescu:2022kof} as the corresponding action with a magnetically charged configuration.
We regard  AMBH with two parameters $w$ and $K$  as NED black holes with power index $s$ and charge term  $\xi(s,q)$.
These NED black holes  include interesting known black holes:  dark matter ($s=3/4,w=1/2$), constant scalar hair ($s=w=1$), charged quantum Oppenheimer-Snyder ($s=3/2,w=2$), Einstein-Euler-Heisenberg ($s=2,w=3$),  and Ned black holes ($s=3,w=5$).
Rotating charged AMBHs can describe rotating charged  NED black holes when replacing anisotropic parameter $w$ and density $K$ by power index $s$ and charge term $\xi(s,q)$.

The paper is organized as follows.
In Sect.~\ref{sec2}, we briefly mention how to find AMBHs from an anisotropic fluid with barotropic equation of state but not an anisotropic matter action.
In Sec.~\ref{sec3}, we obtain NED black holes including known black hole solutions  by considering  a power-law of Maxwell term as an anisotropic matter action.
Sec.~\ref{sec4} is devoted to discussing rotating charged NED black holes and their extremal rotating  NED black holes which are the boundary between rotating charged NED black holes and naked singularity.
Finally, we summarize our results and discuss on relevant issues in Sec.~\ref{sec5}.

\section{AMBHs \label{sec2}}

We wish to mention briefly how to obtain AMBHs.
First of all, we consider the anisotropic energy-momentum tensor
\begin{equation} \label{em-T}
T_\mu^{{\rm AM},\nu}={\rm diag}[-\rho,p_1,p_2,p_3]=\rho \cdot  {\rm diag}[-1,w_1,w_2,w_3],
\end{equation}
whose barotropic  equation of state is described  by
\begin{equation}
p_i=w_i \rho.
\end{equation}
For the anisotropic fluid with  $p_1=-\rho$ ($w_1=-1$) in the radial direction  and $p_2=p_3=w\rho$ ($w_2=w_3=w$) in the angular direction, solving $p_2=-\rho-r\rho'/2$ with $f(r)=1-2m(r)/r$ leads to the energy  density,
the mass function,  and the metric function
\begin{equation} \label{am-c}
\rho=\frac{(1-2w)K}{8\pi r^{2w+2}},\quad m(r)=M+\frac{K}{2r^{2w-1}},\quad  f(r)=1-\frac{2M}{r}-\frac{K}{r^{2w}}.
\end{equation}
One may  introduce $r_o$ as a length dimension  defined by $r^{2w}_o=(1-2w)K$. If one is interested in asymptotically flat geometries, it is  worthwhile to analyze the case of  $w > 1/2$. Then, $K$ is  negative for $w > 1/2$  when the energy density is positive. This case will correspond to $s>\frac{3}{4}$ for $\xi(s,q)<0$.

Then, we consider the Einstein-Maxwell theory with unknown anisotropic matter term ${\cal L}_{\rm AM}$~\cite{Kim:2019hfp,C:2024cnk}
\begin{equation}
I_{\rm CAM}=\int d^4x \sqrt{-g}\Big[\frac{1}{16\pi}(R-F_{\mu\nu}F^{\mu\nu})+{\cal L}_{\rm AM}\Big],
\end{equation}
where $-2\frac{\partial {\mathcal{L}_{\rm AM}}}{\partial g^{\mu\nu}}+\mathcal{L}g_{\mu\nu}$ is assumed to arrive at Eq.(\ref{em-T}).

Its charged black hole solution is given by
\begin{equation}
ds^2_{\rm CAM}=-f(r)dt^2+\frac{dr^2}{f(r)}+r^2(d\theta^2+\sin^2\theta d\phi^2),
\end{equation}
where
\begin{equation}
\tilde{f}(r)=1-\frac{2M}{r}+\frac{q^2}{r^2}-\frac{K}{r^{2w}}. \label{g-fun}
\end{equation}
Here,  ADM mass $M$, charge $q$, and  two parameters $w$ and $K$ may be regarded as hairs for  the charged AMBH.
In this case, one does  not know their origin of $w$ and $K$ because the energy-momentum $T_{\mu\nu}^{\rm AM}$ was used to derive the solution.

\section{NED black holes \label{sec3}}

First of all, we remind the reader that the AMBH solutions (\ref{am-c}) and (\ref{g-fun})  were derived from the anisotropic fluid of $w_1=-1$ with various values of $w_2=w$ as an extension of Reissner-Nordstr\"om (RN) black hole whose energy-momentum tensor is given by
$T^{{\rm RN}~\nu}_{\mu}=\frac{2q^2}{r^4} {\rm diag}[-1,-1,1,1]$.
This indicates  a hint for choosing  $\mathcal{L}_{\rm AM}$ as an extension of the Maxwell term $\mathcal{F}=F_{\mu\nu}F^{\mu\nu}$.

Hence, let us introduce the Einstein-NED action proposed by Hassaine and Martinez~\cite{Hassaine:2008pw}
\begin{equation} \label{HM-A}
I_{\rm HM}=\int d^4x \sqrt{-g}\Big[\frac{R}{16\pi}- \xi\mathcal{F}^s \Big]
\end{equation}
with  an action parameter $\xi$ for the NED term of  $\mathcal{F}$ to the power of $s$.
In this case, the Einstein equation takes the form
\begin{equation}
G_{\mu}^{~\nu}=8\pi  T_{\mu}^{~\nu},
\end{equation}
whose energy-momentum tensor is given by
\begin{equation}
 T_{\mu}^{~\nu}=-\xi\Big(\mathcal{F}^s\delta^\nu_\mu-4s\mathcal{F}^{s-1}F_{\mu\rho}F^{\nu\rho}\Big).
\end{equation}
Its non-linear Maxwell equation is given by
\begin{equation}
\nabla_\mu\Big( \mathcal{F}^{s-1}F^{\mu\nu}\Big)=0.
 \end{equation}
For a magnetically charged case, one finds
\begin{equation}
A_\phi=-q \cos \theta \to F_{\theta\phi}=q \sin\theta \to \mathcal{F}=\frac{2q^2}{r^4}(=2B^2)
\end{equation}
with a magnetic field $B=|\bold{B}|$.
In this case, its energy-momentum tensor  takes the form~\cite{Mazharimousavi:2025lld}
\begin{equation}
T_{\mu}^{~\nu}= \frac{2^s \xi q^{2s}}{r^{4s}} {\rm diag}[-1,-1,2s-1,2s-1],
\end{equation}
which satisfies all (null, weak, strong)  energy conditions for $\xi>0$ and $s\ge\frac{1}{2}$ except the dominant energy condition of $\rho\ge |p_i|$.
The mass function is derived as
\begin{equation}
m(r)=4\pi \int^r r'^2\rho(r')dr'=M+ \frac{ 2^{s+2} \pi \xi q^{2s}}{(3-4s)r^{4s-3}}.
\end{equation}
Then, we obtain the metric function to describe  the NED black hole solution
\begin{equation}
h(r)=1-\frac{2M}{r}-\frac{ \xi(s,q)}{r^{4s-2}},\quad \xi(s,q)=\pi \xi\frac{ 2^{s+3} q^{2s}}{3-4s}. \label{h-fun}
\end{equation}
Importantly, comparing Eq.(\ref{am-c}) with Eq.(\ref{h-fun}) leads to
\begin{equation}
w=2s-1,\quad K=\xi(s,q).
\end{equation}
However, we note that the expression of Eq.(\ref{h-fun}) is valid for $s\not=\frac{3}{4}$.
For $s=\frac{3}{4}$, one finds the metric function~\cite{Li:2012zx,Liang:2023jrj}
\begin{equation}
    h_{s=3/4}(r)=1-\frac{2M}{r}-\frac{q^{3/2}\log[\frac{2^{1/4}r}{q^{3/2}}]}{2^{1/4}r},\quad {\rm for}~\xi=\frac{1}{16\pi}, \label{s3/4}
\end{equation}
which is an asymptotically flat dark matter black hole (DMBH).

Furthermore, it is worthy to note that the NED term for $s=0$ becomes  a constant and the metric tensor  no longer couples dynamically to the gauge field,
even though  we have the Schwarzschild-de Sitter (SdS) black hole.
Although the magnetic configuration $F_{\theta\phi}=q\sin\theta$ is still written down,
it does not contribute any electromagnetic stress-energy tensor. In this sense, SdS black hole is not genuinely a charged solution and the parameter $q$ plays no role.
\begin{figure}[ht]
\centering
\subfigure[$\xi=\frac{1}{16\pi}$ and $q=0.5$]{\includegraphics[width=0.4\textwidth]{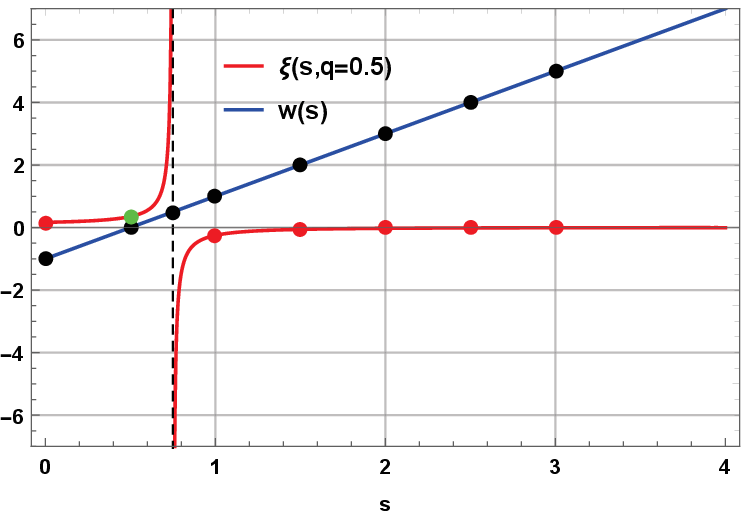}}
\hfill%
\subfigure[$\xi=-\frac{1}{16\pi}$ and $q=0.5$]{\includegraphics[width=0.4\textwidth]{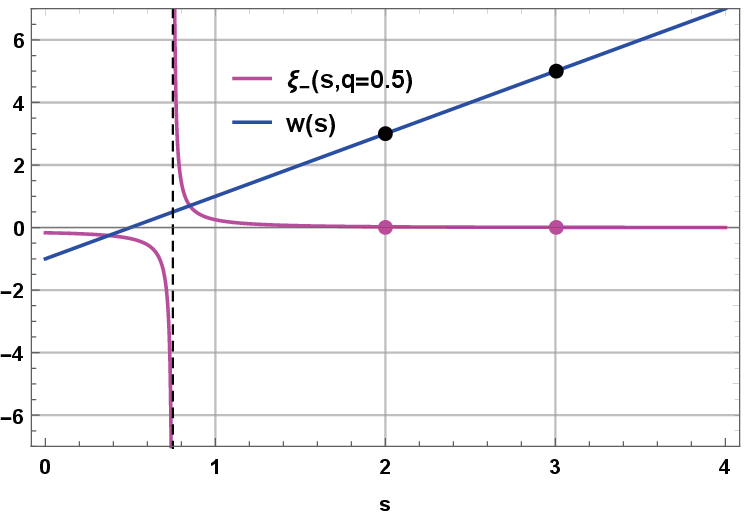}}
\caption{(a) Charge term $\xi(s,q=0.5)$ is as a function of $s$. For $0<s<\frac{3}{4}$, $\xi(s,0.5)$ increases, it blows up at $s=\frac{3}{4}$ [instead, we make a green dot by Eq.(\ref{s3/4})], and it increases negatively for $s>\frac{3}{4}$.
Equation of state parameter $w(s)$ is as a function of $s$. It is negative for $0<s<\frac{1}{2}$, it is zero at $s=\frac{1}{2}$,  and it is positive for $s>\frac{1}{2}$. Eight black dots are displaced for Table~\ref{table}. Six red dots including one green dot (not black hole) are shown for Fig.~\ref{extremal}(a). (b) $\xi_-(s,q=0.5)$ is as a function of $s$. Two magenta dots  represent EEH and Ned black holes. }
\label{xiofs}
\end{figure}

Here, if one chooses the action parameter $\xi=\frac{1}{16\pi}$, there is no ambiguity in describing $\xi(s,q)$ in terms of    charge $q$ and  power-index $s$.
Figure \ref{xiofs} shows the plot of the charge terms $\xi(s,q=0.5)$ and $\xi_-(s,q=0.5)$, as  functions of $s$.  For $0<s<\frac{3}{4}$, $\xi(s,0.5)$ increases, it seems to blow up at $s=\frac{3}{4}$ (apparent feature), and it increases negatively.
In addition, the equation of the state parameter $w(s)$ is a linear function of $s$: negative for $0<s<\frac{1}{2}$; zero at $s=\frac{1}{2}$; positive for $s>\frac{1}{2}$.
For $s=\frac{1}{2}$(its action: $\sqrt{\mathcal{F}})$, its anisotropic term $\xi(s,q)/r^{4s-2}$ becomes constant, while
this term is shown in Eq.(\ref{s3/4}) for $s=\frac{3}{4}$ (its action: $\mathcal{F}^{3/4}$).

\begin{figure}[ht]
\centering
\subfigure[$s=0$]{
\includegraphics[width=0.4\textwidth]{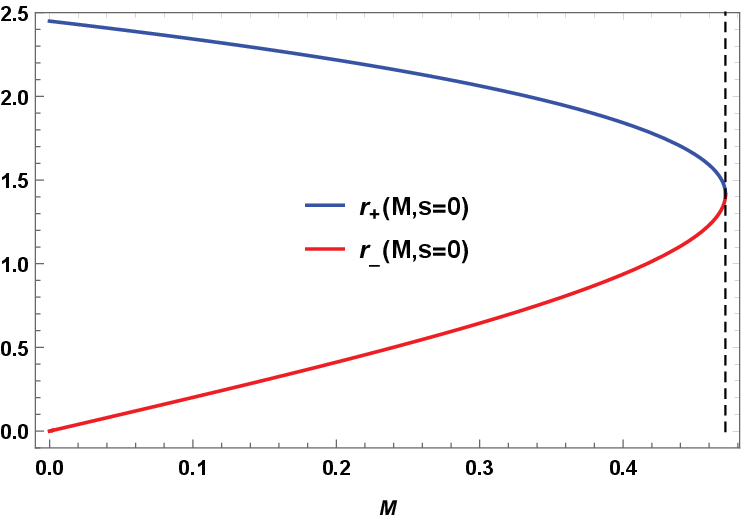}}
\hfill%
\subfigure[$s=1,3/2$]{\includegraphics[width=0.4\textwidth]{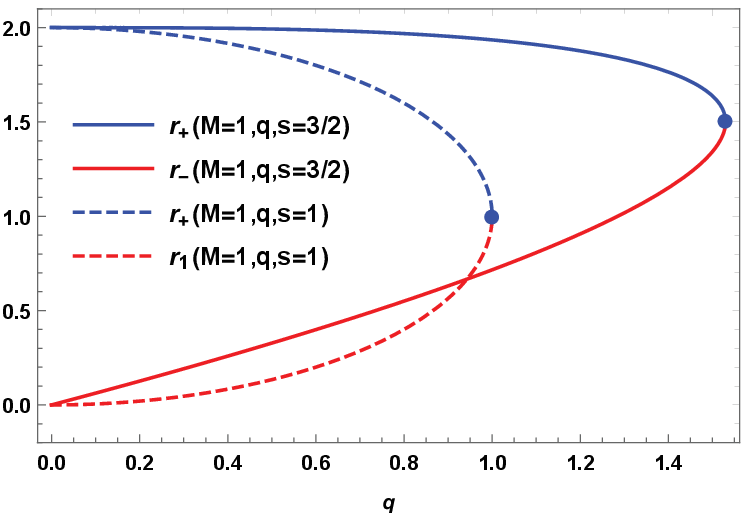}}
\caption{ Horizon structure.  (a) For  $s=0$ and $\xi=\frac{1}{16\pi}$, $r_\pm$ represents the outer/Cauchy horizons of Schwarzschild-de Sitter black hole for $0<M<0.471$. The dashed line is located at extremal point $M_e=0.471$. (b) For $s=1,3/2$,  $\xi=\frac{1}{16\pi}$ and $M=1$. $r_\pm$ represents the outer/inner horizons of RN and cqOS black holes for $0<q<1$ and $0<q<1.53$. Two blue dots  are located at extremal points (1,1) and (1.53.1.5) and they will also  be displaced in Fig.~\ref{extremal}(a). }
\label{fig16}
\end{figure}

This approach recovers famous known black hole solutions (see Table \ref{table}).
For $s=0(w=-1)$, Eq.(\ref{h-fun}) represents Schwarzschild-de Sitter (SdS) black hole~\cite{Brady:1996za}, while  Eq.(\ref{s3/4}) denotes dark matter (DM) black hole for $s=3/4(w=1/2)$. The $s=1(w=1)$ case indicates the RN black hole.
In case of $s=3/2(w=2)$, Eq.(\ref{h-fun}) indicates the charged quantum Oppenheimer-Snyder (cqOS) black hole~\cite{Mazharimousavi:2025lld}, which may become the quantum Oppenheimer-Snyder (qOS) black hole for selecting $P=M$~\cite{Lewandowski:2022zce}.  Now, we would like to mention that the electrically charged black hole solution was derived from the same action (\ref{HM-A})~\cite{Mazharimousavi:2025lld}. Although its form differs slightly from Eq.(\ref{h-fun}), it  can describe the same thing. In this sense, the power-index  $s$ is not considered as a hair of NED black holes, but a sorter of NED black holes. The charge $q$ in $\xi(s,q)$ can be regarded as a hair for NED black holes with mass $M$.
\begin{table}[h]
\center
\begin{tabular}{|c||c|c|c|c|c|c|c|c|}
  \hline
  $s$ &0 &1/2&3/4& 1  & 3/2 & 2  &$\frac{5}{2}$ & 3\\ \hline
  $w(s)$ &$-1$& 0 & 1/2 & 1 & 2 & 3& 4 &5  \\ \hline
   $K/\xi(s,q)$&$\frac{1}{6}$ &$ \frac{q}{\sqrt{2}}$& *& $- q^2/-q_{s}^2$  & $-\frac{\sqrt{2}q^3}{3}$ & $-\frac{2 q^4}{5}$  & $-\frac{2\sqrt{2}q^5}{7}$& $-\frac{4q^6}{9}$\\ \hline
    BH & SdS\cite{Brady:1996za}&N.A. &DM&RN/CSH&cqOS& EEH\cite{Yajima:2000kw}&  &Ned\\ \hline
  RBH & KdS&RG &N.A. &KN\cite{Newman:1965my} && REEH\cite{Breton:2019arv}& &   \\ \hline
\end{tabular}
\caption{Table for classification of known  black holes with respect to $s$ and $w(s)$ for $\xi=\frac{1}{16\pi}$.  (R)BH, SdS, DM\cite{Li:2012zx,Liang:2023jrj}, RN, CSH\cite{Astorino:2013sfa,Myung:2024pob},  cqOS\cite{Mazharimousavi:2025lld}, (R)EEH,  Ned\cite{Myung:2025ecg}, KdS\cite{Akcay:2010vt}, and RG\cite{Rubin:1970zza} denote (rotating) black hole, Schwarzschild-de Sitter, dark matter, Reissner-Nordstr\"om,  constant scalar hair, charged quantum Oppenheimer-Snyder, (Rotating) Einstein-Euler-Heisenberg,  nonlinear electrodynamics, Kerr-de Sitter, and rotating galaxy. Note that $\xi=-\frac{1}{16\pi}$ is chosen for (R)EEH  and Ned black holes. * denotes $\frac{q^{3/2}}{2^{1/4}}\log [2^{1/4}/q^{3/2}]$. }
\label{table}
\end{table}
It should be noted that its horizon structure obtained from $h(r)=0$ depends crucially on the power-index $s$. So, one could not find its horizon structure for general $s$.
As was shown in Fig.~\ref{fig16}, there exist two horizons:
(a) $s=0$, $r_\pm$ represents the outer/Cauchy horizons for SdS black hole for $0<M<0.471$. (b) $s=1,3/2$ and $M=1$. $r_\pm$ represents the outer/inner horizons of RN and cqOS black holes that exist  for $0<q<1$ and $0<q<1.53$. The allowed region $q$ for cqOS is larger than RN. For a given $s$-black hole solution, the  corresponding outer horizon defined by $r_+$ can be used to compute all thermodynamic quantities, scalarization, shadow radius analysis, and greybody and quasinormal modes computations.

Considering the Einstein-Maxwell theory with the NED term
\begin{equation}
I_{\rm EMN}=\int d^4x \sqrt{-g}\Big[\frac{1}{16\pi}(R-F_{\mu\nu}F^{\mu\nu})-\xi \mathcal{F}^s  \Big],
\end{equation}
the charged NED black hole solution is given by
\begin{equation}
\tilde{h}(r)=1-\frac{2M}{r}+\frac{q^2}{r^2}-\frac{\xi(s,q)}{r^{4s-2}}. \label{hc-fun}
\end{equation}
In this case, the mass $M$ and the charge $q$ are  regarded as  hairs, while $s$ is a sorter for charged NED black holes.  We have its two known black hole solutions.
For $s=1(w=1)$, Eq.(\ref{hc-fun}) represents a constant scalar hair (CSH) black hole derived from Einstein-Maxwell-conformally coupled scalar theory~\cite{Astorino:2013sfa,Myung:2024pob}.
In case of $s=2(w=3)$, Eq.(\ref{hc-fun}) indicates  the Einstein-Euler-Heisenberg (EEH) black hole~\cite{Yajima:2000kw} which includes the QED effect of vacuum polarization~\cite{Heisenberg:1936nmg}. The $s=3(w=5)$ case leads to a Ned black hole.
As is shown in Table 1, there are known black hole solutions derived from the Einstein-NED action.

\section{ Rotating charged NED  black holes \label{sec4}}

The rotating charged black hole solution with NED matter  is obtained by applying the Newman-Janis algorithm to a static spherically symmetric solution Eq.(\ref{hc-fun}) as~\cite{Boyer:1966qh, Carter:1968ks, Kim:2019hfp}
\begin{eqnarray}
\label{metric}
ds^2&=&  - F(r, \theta) dt^2 -2[1-F(r, \theta)]a\sin^2\theta dt d\phi + \frac{\Sigma}{\rho^2} \sin^2\theta d\phi^2 + \frac{\rho^2}{\triangle} dr^2 + \rho^2 d\theta^2 \, \nonumber \\
&=& - \frac{\triangle}{\rho^2} (dt - a\sin^2\theta d\phi)^2 + \frac{\sin^2\theta}{\rho^2}[a dt -(r^2+a^2) d\phi]^2  +  \frac{\rho^2}{\triangle} dr^2 + \rho^2 d\theta^2 \,,
\end{eqnarray}
 where
\begin{eqnarray}
 F(r, \theta)&=&1-\frac{2Mr-q^2+\xi(s,q)r^{4(1-s)}}{\rho^2} ,\quad a=\frac{J}{M},\quad\rho^2=r^2+ a^2\cos^2\theta, \nonumber \\
 \triangle & =& r^2 + a^2 + q^2 -2Mr - \xi(s,q) r^{4(1-s)}, \nonumber \\
 \Sigma&=& \rho^2(r^2+a^2) + [2Mr -q^2 + \xi(s,q) r^{4(1-s)}]a^2\sin^2\theta. \label{F-rho}
\end{eqnarray}
The  $\xi(s,q)=0$ case leads to the KN black hole, and $\xi(s,q)=0$ with $q=0$ corresponds to the Kerr black hole~\cite{Kerr:1963ud}, regarding as two references to the rotating black hole. At this stage, we mention that the event (outer)  horizon ($r_+$) for the spacetime (\ref{metric}) is located at the largest radius as a solution to $\triangle=0$~\cite{Carter:1969zz}, leading to the event horizon for the KN black hole with  $\xi(s,q)=0$.   The region between the event horizon and the infinite-redshift surface  of  $F(r, \theta) =0$  is called the ergosphere~\cite{Ruffini:1970sp}.
\begin{figure}[ht]
\centering
  \subfigure[ $M=1,a=0$]{\includegraphics[width=0.3\textwidth]{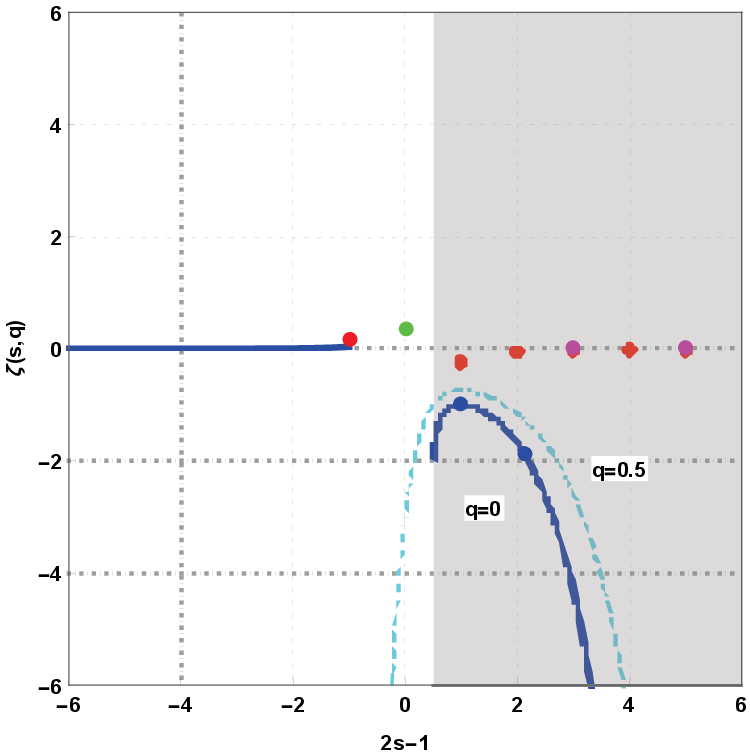}}
\hfill%
\subfigure [$M=1,a\not=0$]{\includegraphics[width=0.3\textwidth]{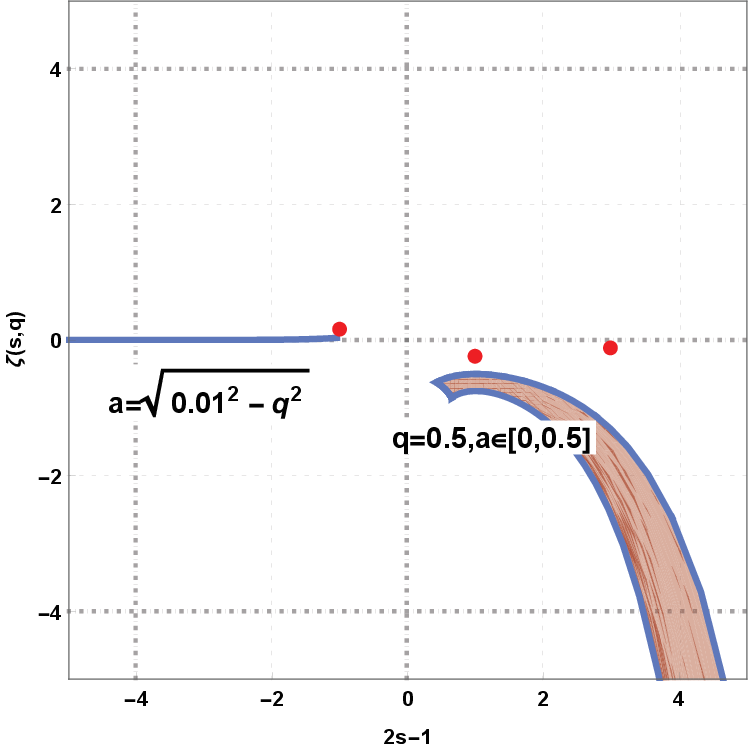}}
\hfill%
\subfigure[ $M=1$]{\includegraphics[width=0.3\textwidth]{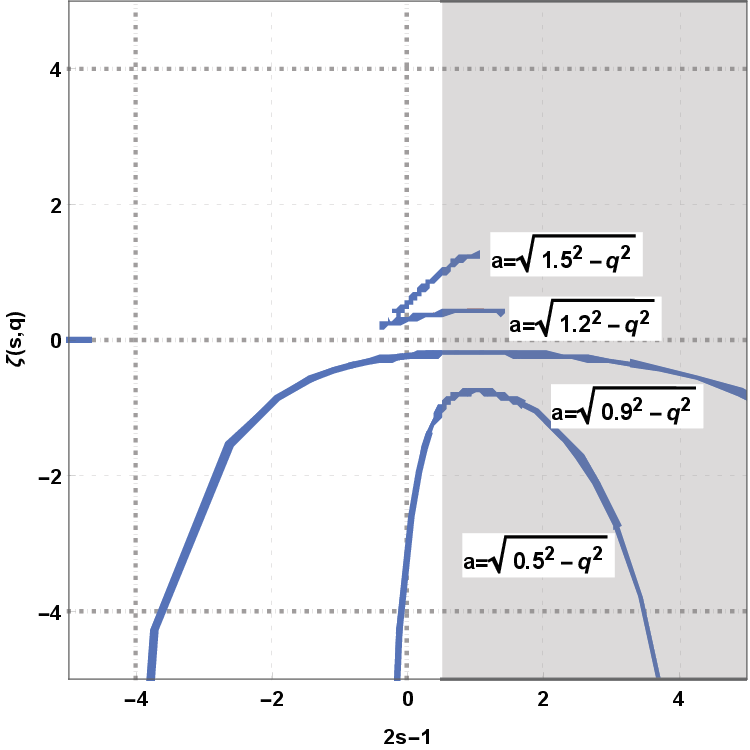}}
\caption{(a) Curves for extremal NED black holes with $q=0,q=0.5$ and $a=0$. All  dots including two magenta dots shown in Figs.~\ref{xiofs}(a) and \ref{xiofs}(b) are located above these extremal curves. The shaded region of asymptotically flat black holes denotes the region of $s\ge\frac{3}{4}(w\ge1/2)$. (b) Curves for extremal rotating NED black holes with $q=0,0.5$ and $a\not=0$.  (c) Four curves of extremal rotating  NED black holes : $a(q)=\sqrt{0.5^2-q^2},~\sqrt{0.9^2-q^2},~\sqrt{1.2^2-q^2}$, and $\sqrt{1.5^2-q^2}$. The shaded region indicates for $s\ge\frac{3}{4}$. Each solid curve separating the black hole (above curve) from naked singularity (below curve) are parametrically described by Eq.(\ref{f-curves}).  }
\label{extremal}
\end{figure}
In this case, its horizon structure critically depends  on $s>0$ and $\xi$, so it is difficult to display  it.
However, requiring $\Delta=0$ and $\Delta'=0$ simultaneously, the extremally rotating black hole whose horizons are degenerate  is  determined  by~\cite{Badia:2020pnh}
\begin{equation} \label{f-curves}
    2s-1=\frac{a^2+q^2-Mr}{\Delta_{\rm KN}},\quad  \xi(s,q)=\frac{\Delta_{\rm KN}}{r^{\frac{2r(r-M)}{\Delta_{\rm KN}}}}
\end{equation}
with $\Delta_{\rm KN}=r^2-2Mr +a^2+q^2$. This extremal rotating black hole represents the boundary between rotating charged NED black hole and naked singularity.

As an example, we consider the extremal NED black holes.  Figure \ref{extremal} shows the curves for the extremal NED black hole, where $w$ is given by $2s-1$.
The solid curve in Fig.~\ref{extremal}(a) includes two blue dots for extremal RN BH at (1,$-$1) and extremal cqOS BH at (2.13,$-1$.88) shown in Fig.~\ref{fig16}(b). This is given  by $2s-1=-Mr/\Delta_S$ and $\xi(s,q)
=\Delta_S/r^{2r(r-M)/\Delta_S}$ with $\Delta_S=r^2-2Mr$.   The extremally charged NED
black holes with $q=0.5$ (dashed curve in Fig.~\ref{extremal}(a)) are determined by $2s-1=(q^2-Mr)/\Delta_{\rm RN}$ and $\xi(s,q)
=\Delta_{\rm RN}/r^{2r(r-M)/\Delta_{\rm RN}}$ with $\Delta_{\rm RN}=r^2-2Mr+q^2$.
Their extremal black hole curves are shown in Fig.~\ref{extremal}(a). All  dots, including  two green dots shown in Fig.~\ref{xiofs}(a), are also located above the extremal curves in Fig.~\ref{extremal}(a). The first red dot denotes SdS. The first green dot represents an object but not a black hole because there is no corresponding extremal curve, while the second green dot denotes DM black hole. All six dots located inside the shaded region  represent as asymptotically flat NED black holes shown in Fig.~\ref{xiofs}(a).
At this stage, we wish to point out that two dots at $w(=2s-1)=3,5$ include two red dots of $\xi(s,q)=-0.025,-0.0069$ (Fig. 1(a)) for NED black holes and two magenta dots of $\xi_(s,q)=0.025,0.0069$ (Fig.~\ref{xiofs}(b)) for  charged NED black holes (EEH and Ned).
This implies that $s(w)$ on the $x$-axis classifies NED black holes, while $\xi(s,q)(K)$ on the $y$-axis denotes the corresponding charge term.

For extremal rotating black holes with $M=1,q=0,0.5$, see  Fig.~\ref{extremal}(b), which includes
three red dots for KdS, KN, and (R)EEH black holes. All are located above the extremal curves.
However, plotting the curves parametrically when choosing $a(q)$ leads to four curves  where there are extremal rotating  NED black holes, as shown in Fig.~\ref{extremal}(c). Unlike the extremal KN spacetime with $q^2+a^2\le M^2(=1)$ (below two negative  curves),  extremal rotating  NED black holes can allow for $q^2+a^2> M^2(=1)$ (above two positive curves). In addition, the existence $s$-region for the former is larger than that for the latter.  We note that each solid curve separates rotating charged NED black hole (above curve) from a naked singularity (below curve).

In addition, we may read off the orthonormal frame from Carter's form of Eq.~(\ref{metric}). The corresponding covariant tetrad is given by
\begin{eqnarray}
\label{cotetrad}
&&e^{\hat t}_{\mu} = \frac{\sqrt{\triangle}}{\rho} (1,0,0,-a\sin^2\theta)\,, \quad
e^{\hat \phi}_{\mu} =  \frac{\sin\theta}{\rho} (a,0,0, -(r^2+a^2)) \,, \nonumber \\
&& e^{\hat r}_{\mu} = \frac{\rho}{\sqrt{\triangle}} (0,1,0,0) \,, \quad
  e^{\hat \theta}_{\mu} = \rho (0,0,1,0) \,
\end{eqnarray}
and the contravariant tetrad takes the form~\cite{Carter:1968ks, Azreg-Ainou:2014nra}
\begin{eqnarray}
\label{otetrad}
&&e^{\mu}_{\hat{t}}= \frac{1}{\rho\sqrt{\triangle}}(r^2+a^2,0,0,a)\,,\quad e^{\mu}_{\hat{\phi}} =-\frac{1}{\rho\sin\theta}(a\sin^2\theta,0,0,1)
 \,, \nonumber \\
&& e^{\mu}_{\hat{r}}= \frac{\sqrt{\triangle}}{\rho}(0,1,0,0)\,, \quad e^{\mu}_{\hat{\theta}}=\frac{1}{\rho}(0,0,1,0)\,.
\end{eqnarray}
Using Eq.(\ref{otetrad}), an observer located at this orthonormal frame can read the physical quantities of energy density and pressures as
\begin{eqnarray}
\label{varepsilon}
&&\varepsilon(=-p_{\hat{r}})=\frac{G_{\mu\nu}e^{\mu}_{\hat{t}}e^{\nu}_{\hat{t}}}{8\pi}=\varepsilon_e + \varepsilon_{\rm AM} =\frac{1}{8\pi \rho^4}\Big[q^2 +2^{s+3} \pi \xi q^{2s}  r^{4(1-s)}\Big]\,,\label{ep-1} \\
&&p_{\hat{\theta}}(=p_{\hat{\phi}})= \frac{G_{\mu\nu}e^{\mu}_{\hat{\theta}}e^{\nu}_{\hat{\theta}}}{8\pi}
 = \frac{[(s+1)q^2\rho^2 +2^{s+2} \pi \xi q^{2s}  r^{4(1-s)}] [(2s-1)\rho^{2}-a^2\cos^2\theta]}{4\pi r^2 \rho^4}\,. \label{ep-2}
\end{eqnarray}

In the non-rotating limit of $a\to 0$, one obtains  the static energy density and pressure  from Eqs.(\ref{ep-1}) and (\ref{ep-2})
\begin{equation}
\varepsilon_{s}=\frac{q^2}{8\pi r^4}+\frac{2^s \xi q^{2s}}{r^{4s}}, \quad p^s_{\hat{\theta}}=\frac{(s+1)q^2}{4\pi r^4}+\frac{(2s-1)2^s\xi q^{2s}}{r^{4s}}.
\end{equation}
It is important to note that the parameters $\xi(s,q)$ and $s$  control the charge term
and anisotropy of the NED matter.

As shown in Table~\ref{table} (Fig.~\ref{xiofs}(b)), there exist three rotating NED black holes for $s=0,1,2$ ($w=-1,1,3)$.
Hence, one expects to read off  a lot of  rotating charged  NED black holes from Fig.~\ref{extremal}(c).
Finally, we would like to mention that for the $s=1/2(w=0)$ case,  its Ref.~\cite{Rubin:1970zza} in Table~\ref{table} is an observational paper on the galactic rotation curve, but not a field theory derivation.

\section{Conclusion \label{sec5}}

Although research on anisotropic fluid black holes is ongoing, it has been difficult to construct black hole solutions starting from an action. This is primarily due to the absence of a recognized action for such  a fluid matter. In this study, we have introduced a nonlinear electrodynamics (NED) term to construct black hole solutions starting from the corresponding explicit action.

We have found a correspondence between   anisotropic matter black holes (AMBHs) with anisotropic  parameter $w$ and energy density $K$ and  nonlinear electrodynamics (NED) black holes with power-index $s$ and   $\xi(s,q)$.  This implies that AMBHs could be regarded as  NED black holes. Here, $s$ is identified with  a classifier  of NED black holes, while $\xi(s,q)$ corresponds to the charge term that includes charge $q$ as a hair.
These NED black holes have included  dark matter ($w=1/2)$, Reissner-Nordstr\"om and  constant scalar hair ($w=1$), charged quantum Oppenheimer-Snyder ($w=2$),  Einstein-Euler-Heisenberg ($w=3$), and Ned ($w=5$)  black holes derived from their known actions.

Finally, we would like to mention that  rotating charged AMBHs can describe rotating charged NED black holes as well when replacing $w$ and $K$ by $s$ and $\xi(s,q)$.
The extremal rotating NED curves that are the boundary between the rotating charged NED black hole and the naked singularity were derived as functions of the rotation parameter $a(q)$.

\section*{Acknowledgments}

W.~L (RS-2026-25484780), Y.~S.~M (RS-2022-NR069013), and Center for Quantum Spacetime (CQUeST) of Sogang University (RS-2020-NR049598) were supported by Basic
Science Research Program through the National Research Foundation of Korea funded by the Ministry of Education.
We are grateful to Hyeong-Chan Kim for his hospitality during our visit to the Tangeumdae Workshop.

\end{document}